\documentclass[aps,floats,twocolumn]{revtex4}
\usepackage{tabularx,graphicx}
\usepackage{epsfig}
\usepackage{amsmath}
\usepackage{bbm}
\usepackage{graphicx}

\begin{document}

\title{Kohn-Luttinger superconductivity in graphene}
\author{J. Gonz\'alez  \\}
\address{Instituto de Estructura de la Materia,
        Consejo Superior de Investigaciones Cient\'{\i}ficas, Serrano 123,
        28006 Madrid, Spain}

\date{\today}

\begin{abstract}

We investigate the development of superconductivity in graphene when the 
Fermi level becomes close to one of the Van Hove singularities of the electron
system. The origin of the pairing instability lies in the strong anisotropy
of the $e$-$e$ scattering at the Van Hove filling, which leads to a channel
with attractive coupling when making the projection of the BCS vertex on the 
symmetry modes with nontrivial angular dependence along the Fermi line. We show
that the scale of the superconducting instability may be pushed up to 
temperatures larger than 10 K, depending on the ability to tune the system
to the proximity of the Van Hove singularity.

\end{abstract}

\maketitle

Since the fabrication in 2004 of single atomic layers of carbon, this new
material (so-called graphene) has been attracting a lot of attention\cite{att}. 
The undoped system has conical valence and conduction bands meeting at two
different Fermi points (known as Dirac points)\cite{geim,kim}. This 
peculiar dispersion has shown to be at the origin of a number of remarkable 
effects, like the existence of a minimum conductivity at the charge neutrality 
point\cite{paco,kat,twor,mac}.

From the point of view of possible applications, the interest in graphene 
has been driven by the large electron mobilities attained in typical experimental
samples. Another remarkable property is that graphene can be used to build 
Josephson junctions when placed between superconducting contacts\cite{delft}. 
It becomes then quite intriguing whether graphene may 
support superconducting correlations on its own under suitable experimental 
conditions. On theoretical grounds, it is known that a model based on the 
conical dispersion requires a minimum strength of the pairing interaction for
the development of a superconducting instability in the undoped system\cite{mar}. 
There have been already several proposals to drive graphene towards a pairing 
instability upon doping, 
placing the emphasis on the role of topological defects\cite{nos}, the effect of a 
metallic coating\cite{uch}, or the possibility of inducing superconductivity by 
electronic correlations\cite{don,jk}.

In this paper we investigate a different route to superconductivity in 
graphene, when the Fermi level is close to one of the Van Hove singularities 
(VHSs) of the electron system. These are points characterized by a divergent 
density of states, which has the effect of enhancing the magnetic and 
superconducting correlations\cite{mark}. 
The origin of the pairing instability lies in the strong anisotropy of the 
Fermi line at the Van Hove filling, and it proceeds following in essence the 
same mechanism proposed by Kohn and Luttinger to show that superconductivity 
can arise out of purely repulsive interactions\cite{kl}. This is possible as 
long as the $e$-$e$ scattering becomes highly anisotropic, so that a channel 
with attractive coupling may appear when making the projection on the symmetry 
modes with nontrivial angular dependence over the Fermi surface.

\begin{figure}
\begin{center}
\mbox{\epsfxsize 6.5cm \epsfbox{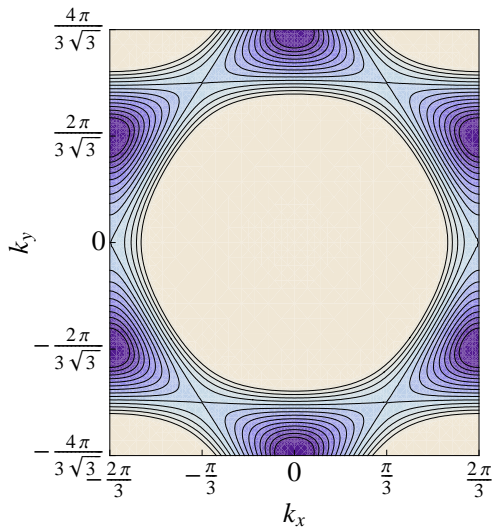}}
\end{center}
\caption{Plot of energy contour lines around the saddle points of the valence
band of graphene, obtained from a tight-binding model with first, second, and
third-neighbor hopping parameters given by Eqs. (\ref{tp})-(\ref{d}) for 
$s = 0.1$ .}
\label{one}
\end{figure}

In the case of graphene, there are two VHSs located at 
about 3 eV above and below the Dirac points in the spectrum. Each of the 
singularities correspond to the existence of three inequivalent saddle points 
of the dispersion at the boundary of the Brillouin zone, as shown in Fig. 
\ref{one}. In order to find the dominant electronic instability arising 
from the divergent density of states, one has first to determine the shape of 
the Fermi line when the Fermi level is close to the VHS. For this purpose, 
we have characterized the energy contour lines around the saddle points of the
valence band by means of a tight-binding model, suited to fit the dispersion 
$\varepsilon ({\bf k})$ known from angle-resolved photoemission spectroscopy 
(ARPES)\cite{ohta}. In the model, we have considered the transfer integrals 
for first, second, and third neighbors of the graphene lattice, 
labelled respectively by $t$, $d$ and $t'$, and the overlap integral $s$ 
between first neighbors. 
In terms of these parameters, the Fermi velocity at the Dirac points is given 
by $v_F = (3/2)(t - 2t' + 3sd)$, which must be set to the value found in 
graphene, $v_F \approx 2.7$ eV. Moreover, the level of the saddle points 
relative to the Dirac points turns out to be $3d + (t - 3t' -2d)/(1+s)$, 
which must correspond to the value $\approx 2.7$ eV found in ARPES \cite{ohta}. 
With this input, we arrive at the two conditions
\begin{eqnarray}
t'  & \approx &  d - 2.7 s                           \label{tp}               \\
t  & \approx &  2.7 + 2d - 5.4 s -3sd 
\label{t}
\end{eqnarray}
Finally, we can adjust the parameters to reproduce the curvature of the 
dispersion at the saddle points\cite{ohta}, arriving at a dependence of the 
hopping $d$ which is linear on $s$ to very high accuracy:
\begin{equation} 
d \approx 0.07 + 2.8 s + O(s^2)                           \label{d}
\end{equation}

The ARPES dispersion around the saddle points can then be fitted leaving free  
the overlap integral $s$. The important point is that the hopping parameter 
$t'$ remains always constrained to a very small value, $t' \approx 0.1$. This
parameter controls the approximate nesting of the Fermi line, that is, the 
possibility of having regions in which the dispersion satisfies 
$\varepsilon ({\bf k}) \approx - \varepsilon ({\bf Q} + {\bf k})$, with a 
fixed momentum ${\bf Q}$. This is realized in the model with hopping 
parameters obtained from Eqs. (\ref{tp})-(\ref{d}), as observed in a typical 
plot of energy contour lines shown in Fig. \ref{one}. The measure of the 
nesting is given by the susceptibility $\chi ({\bf Q}, \omega )$ at the 
momentum ${\bf Q}$ connecting two inequivalent saddle points, which diverges 
in any event due to the singular density of states. We may approximate for
instance the dispersion with the deviation $\delta {\bf k}$ from 
the saddle point $(2 \pi /3 , 0)$ by 
\begin{equation}
\varepsilon ({\bf k}) \approx 
                       - \alpha \: \delta k_x^2 + \beta \: \delta k_y^2 
\label{sp}
\end{equation}
Then we have
\begin{equation}
\chi ({\bf Q}, \omega ) \approx \frac{1}{2 \sqrt{3} \pi^2} 
     \frac{c'}{\alpha + \beta}  \log (\omega)
\label{sQ}
\end{equation}
where the nesting instability at $\alpha = 3 \beta $ appears in the 
dependence 
\begin{equation}
c' = \log \left(\frac{1+\sqrt{\beta /3\alpha }}{1-\sqrt{\beta /3\alpha }}\right)
   + \log \left(\frac{1+\sqrt{3\beta /\alpha }}{1-\sqrt{3\beta /\alpha }}\right)
\end{equation}
This has to be confronted with the susceptibility at vanishing momentum transfer,
which is 
\begin{equation}
\chi ({\bf 0}, \omega ) \approx \frac{1}{4 \pi^2} 
     \frac{1}{\sqrt{\alpha \beta }}  \log (\omega)
\label{s0}
\end{equation}
For the lower VHS in graphene, sensible values of the parameters are 
$\alpha \approx 7.41$ and $\beta \approx 2.03$ (as obtained for 
$s = 0.1$), which give $c' \approx 3.6$. We see therefore that the 
susceptibility at momentum ${\bf Q}$ prevails over that at vanishing momentum. 

When the Fermi line becomes close to the saddle points, the large anisotropy 
of the $e$-$e$ scattering may actually induce a pairing instability. This
can be understood within a renormalization group framework, by reelaborating 
the argument given by Kohn and Luttinger long time ago\cite{kl}. We will denote 
the interaction vertex by $V (\theta, \theta' )$ for the particular case of BCS 
kinematics, in which the incoming (outgoing) particles collide with zero total 
spin and zero total momentum and the angle $\theta $ ($\theta' $) locates the 
position of the spin-up particle over the Fermi line. The BCS vertex gets 
corrections at low energies by the effect of the high-energy electron modes 
in slices between energy $\Lambda $ and $\Lambda + d\Lambda $ about the Fermi
level\cite{sh}. The integration of these modes gives the variation
\begin{equation}
d V (\theta, \theta' ) = \frac{d\Lambda }{\Lambda } 
   \int_0^{2\pi } \frac{d\theta'' }{(2\pi )^2} 
    \frac{\partial k_{\parallel}}{\partial \theta''} \frac{1}{v (\theta'')}
     V (\theta, \theta'') V (\theta'', \theta' )  
\end{equation}
where $v (\theta'') $ is the gradient of the dispersion and 
$\partial k_{\parallel}/\partial \theta'' $ is the variation of the momentum
along the slice parametrized by the angle $\theta'' $. 
We can write the above equation in more compact form by passing to the 
variable 
\begin{equation}
\phi (\theta ) = \frac{1}{2\pi n(\Lambda )} \int_0^{\theta }  
   \frac{d\theta'' }{v (\theta'' )} 
            \frac{\partial k_{\parallel}}{\partial \theta'' }
\end{equation}
where the density of states $n(\Lambda )$ is introduced so that the new 
variable also ranges from 0 to $2 \pi $ \cite{zs}. After defining the 
transformed vertex by $\widetilde{V} (\phi, \phi' ) = V (\theta, \theta' )$, 
we get
\begin{equation}
\frac{\partial \widetilde{V} (\phi, \phi' )}{\partial \log \Lambda } = 
 \frac{n(\Lambda )}{2\pi } \int_0^{2\pi } d\phi'' \widetilde{V} (\phi, \phi'' ) 
      \widetilde{V} (\phi'', \phi' )
\label{scaling}
\end{equation}
We can further decompose the vertex $\widetilde{V} (\phi, \phi' )$ in terms of 
the eigenmodes $\Psi_m^{\gamma } (\phi)$ for the different representations 
$\gamma $ of the point symmetry group,
\begin{equation}
\widetilde{V} (\phi, \phi' ) = \sum_{\gamma, m, n} V_{m,n}^{\gamma }
       \Psi_m^{\gamma } (\phi)   \Psi_n^{\gamma } (\phi')
\label{dec}
\end{equation}
We obtain then the set of coupled scaling equations 
\begin{equation}
\frac{\partial V_{m,n}^{\gamma }}{\partial \log \Lambda } =
 n(\Lambda ) \: 
  \sum_{s}  V_{m,s}^{\gamma }  V_{s,n}^{\gamma }
\label{rg}
\end{equation}
In this framework, we recover the analogue of the Kohn-Luttinger mechanism 
when any of the couplings $V_{m,n}^{\gamma }$ turns out to be 
negative, in such a way that an unstable flow develops when the Fermi line is
approached in the low-energy limit $\Lambda \rightarrow 0$.

\begin{figure}
\begin{center}
\mbox{\epsfxsize 6.5cm \epsfbox{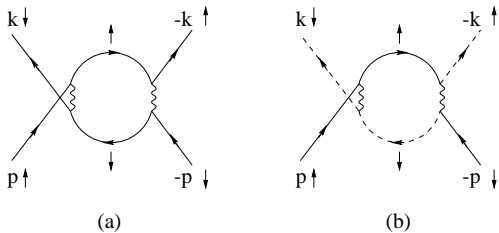}}
\end{center}
\caption{Lowest-order particle-hole corrections to the BCS vertex for a bare
on-site interaction. Full and dashed lines represent the propagation of 
electrons around different saddle points.}
\label{two}
\end{figure}

We remark that the scaling equation (\ref{scaling}) encodes the corrections 
to the BCS vertex that are logarithmically divergent at low energies in the 
particle-particle channel. If we start solving the scaling equation at 
an intermediate energy scale $\Lambda $, the initial values of the function 
$\widetilde{V} (\phi, \phi')$ will be dictated by the bare interaction 
as well as by {\em regular} corrections to it, given in general by finite 
diagrams in the particle-hole channel. Previous scaling analyses of electrons 
near a VHS have shown that only the momentum-independent component of the 
interaction potential is not irrelevant at low energies\cite{np}, which is 
consistent with the large screening effects from the divergent density of 
states. For this reason, we may consider that a bare on-site repulsion $U$
between electron densities with opposite spin provides a sensible form of 
interaction close to the Van Hove filling. The relevant point is that the 
bare short-range interaction gives rise to particle-hole corrections to the 
BCS vertex like those represented in Fig. \ref{two}. If we measure the angles 
$\phi , \phi' $ with respect to the $x$-axis, we observe for instance 
that the value of $\widetilde{V} (0, \pi /3)$ will be enhanced by the 
particle-hole susceptibility (\ref{sQ}) at momentum ${\bf Q}$, while 
$\widetilde{V} (0, 0)$ is enhanced by the particle-hole susceptibility 
(\ref{s0}) at zero momentum. The prevalence of $\chi ({\bf Q}, \Lambda )$
implies that 
\begin{equation}
\widetilde{V} (0, 0) - \widetilde{V} (0, \pi /3 ) 
  \sim 3 U^2 \chi ({\bf 0}, \Lambda ) - 2 U^2 \chi ({\bf Q}, \Lambda ) < 0
\label{cond}
\end{equation}
In general, we can anticipate the dominant terms in the modulation of the 
BCS vertex complying with the symmetry of the Fermi line:
\begin{eqnarray}
\lefteqn{\widetilde{V} (\phi , \phi )  \approx    c_0 + c_6 \cos (6 \phi ) 
                         + \ldots}      \label{mod1}                      \\               
\lefteqn{\widetilde{V} (0 , \phi )  \approx    c_0' + c_2' \cos (2 \phi )
              + c_4' \cos (4 \phi ) + c_6' \cos (6 \phi ) + \ldots }    \\       
 & & \widetilde{V} (\phi , \pi - \phi )  \approx    c_0'' 
  + c_2'' \cos (2 \phi ) + c_4'' \cos (4 \phi ) \;\;\;\;\;\;\;\;\;\;\;\;\;     
                                 \;\;\;\;\;\;\;\;      \nonumber         \\ 
 & &  \;\;\;\;\;\;\;\;\;\;\;\;\;\;\;\;\;\;\;\;\;\;\;\; 
            + c_6'' \cos (6 \phi )   +  \ldots
\label{mod3}   
\end{eqnarray}
As we will see, several relations can be obtained between the coefficients in 
(\ref{mod1})-(\ref{mod3}) by estimating the strength of the different 
scattering processes around the Fermi line.

The Fourier expansions (\ref{mod1})-(\ref{mod3}) match well with the 
decomposition (\ref{dec}) of the BCS vertex in terms of the basis functions
$\Psi_n^{\gamma } (\phi)$. The point symmetry group is $C_{6v}$, which 
has 6 irreducible representations. Four of them are one-dimensional, with 
respective sets of basis functions given by $\{ \cos (6n\phi) \}$, 
$\{ \sin (6n\phi) \}$, $\{ \cos ((6n+3)\phi) \}$, $\{ \sin ((6n+3)\phi) \}$ 
($n$ being always an integer). The other two representations are 
two-dimensional and have sets of basis functions which can be represented by 
$\{ \cos (m \phi) , \sin (m \phi) \}$, with the integer $m$ running over 
all values that are not multiple of 3 and which are odd for one of the 
representations and even for the other. We can write therefore an expansion
of the BCS vertex following (\ref{dec}) and matching the modulations in 
(\ref{mod1})-(\ref{mod3}):
\begin{eqnarray}
\lefteqn{\widetilde{V} (\phi, \phi' )    =   V_{0,0} 
   + 2 V_{0,6} (\cos (6\phi ) + \cos (6\phi' ))}                 \nonumber    \\
 & & +  2 V_{2,2} (\cos (2\phi ) \cos (2\phi') + \sin (2\phi ) \sin (2\phi')) 
                                                                \nonumber    \\
 & & +  2 V_{2,4} (\cos (2\phi ) \cos (4\phi') - \sin (2\phi ) \sin (4\phi') + 
             \phi \leftrightarrow \phi' ) \;\;\;\;\;\;          \nonumber     \\
 & & +  2 V_{3,3} \cos (3\phi ) \cos (3\phi')  
         +  2 V_{3,3}' \sin (3\phi ) \sin (3\phi')  +   \ldots    
\label{exp}
\end{eqnarray}

One can check that the terms displayed in (\ref{exp}) account for the
dependence of the BCS vertex in Eqs. (\ref{mod1})-(\ref{mod3}). We identify
actually the different coefficients 
$c_2' = 2V_{2,2} + 2V_{2,4}, c_4' = 2V_{2,4},c_6' = 2V_{0,6}$, 
on the one hand, and $c_2'' = 4 V_{2,4}, c_4'' = 2V_{2,2},
c_6'' = 4 V_{0,6} + V_{3,3}' $, on the other hand. The constraint 
(\ref{cond}) can be translated then to these couplings, since 
$\widetilde{V} (0, 0) - \widetilde{V} (0, \pi /3 )$ is given by the 
combination $3(c_2' + c_4')/2 = 3(c_2'' + c_4'')/2$. We find
\begin{equation}
 3 (V_{2,2} + 2 V_{2,4}) 
\sim 3 U^2 \chi ({\bf 0}, \Lambda ) - 2 U^2 \chi ({\bf Q}, \Lambda )
\label{cond2}
\end{equation}
This already points at the existence of an unstable flow in the channel
corresponding to the representation with $d$-wave symmetry as long as
$\chi ({\bf 0}, \Lambda ) < \chi ({\bf Q}, \Lambda )$. A closer
inspection reveals actually that both couplings $V_{2,2}$ and $V_{2,4}$
must be negative. To show this, one more constraint can be 
enforced by noticing that  
$\widetilde{V} (\pi /6, -\pi /6 ) - \widetilde{V} (\pi /2, -\pi /2 )
= 3(c_2'' - c_4'')/2 $. The relevant point is that the corrections to
the BCS vertex at those angles are not singular at low energy $\Lambda $,
as they involve scattering by particle-hole processes that scale at most 
as $\sim \sqrt{\Lambda }$ \cite{inst}. Therefore, the dominant contribution
to both couplings $V_{2,2}$ and $V_{2,4}$ arises from the logarithmic
dependence on $\Lambda $ at the right-hand-side of Eq. (\ref{cond2}).

As long as $V_{2,2}$ and $V_{2,4}$ are negative, we have a pairing instability 
in the system, whose critical scale can be estimated by solving the coupled 
scaling equations (\ref{rg}) in the relevant symmetry channel. We can truncate
the set of equations to 
\begin{equation}
\frac{\partial }{\partial \log \Lambda } 
\left(
\begin{array}{cc}
V_{2,2}  &   V_{2,4}    \\
V_{4,2}    &   V_{4,4}
\end{array}         \right)
\approx n (\Lambda ) 
\left(
\begin{array}{cc}
V_{2,2}  &   V_{2,4}    \\
V_{4,2}    &   V_{4,4}
\end{array}         \right)
\left(
\begin{array}{cc}
V_{2,2}  &   V_{2,4}    \\
V_{4,2}    &   V_{4,4}
\end{array}         \right)
\label{mrg}
\end{equation}
Eq. (\ref{mrg}) can be easily solved by passing to the eigenvalues 
$\lambda_1, \lambda_2$ of the matrix of couplings. The scaling equations
read for them
\begin{equation}
\frac{\partial }{\partial \log \Lambda} \lambda_j = n(\Lambda ) \lambda_j^2
\label{erg}
\end{equation}
We remark that $V_{4,4}$
does not appear at the dominant level in the expansion of the BCS vertex, 
implying that such a coupling must be much smaller than those displayed in 
(\ref{exp}). The two eigenvalues can be approximated then by 
$\lambda_{1,2} = (V_{2,2} \pm \sqrt{V_{2,2}^2 + 4V_{2,4}^2})/2$ .
It is clear that the positive eigenvalue $\lambda_1$ will vanish in the 
low-energy limit, while the signature of the pairing instability will be given 
by the growth of $\lambda_2 $ towards very large negative values as 
$\Lambda \rightarrow 0$. 

\begin{figure}
\begin{center}
\mbox{\epsfxsize 5.5cm \epsfbox{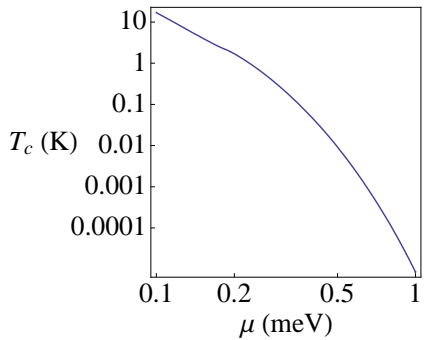}}
\end{center}
\caption{Plot of the temperature $T_c $ obtained by identifying $k_B T_c$ with
the energy at which the eigenvalue $\lambda_2 $ diverges, as a function of the 
deviation $\mu $ of the VHS from the Fermi level.}
\label{three}
\end{figure}

To find the behavior of the negative eigenvalue, we take for $n(\Lambda )$ 
the density of states about a VHS at an energy $\mu $ from the Fermi level:
\begin{equation}
n (\varepsilon ) \approx \frac{3}{4\pi^2} \frac{1}{\sqrt{\alpha \beta }} 
   \log \left( \frac{\Lambda_0}{|\varepsilon - \mu |} \right)
\end{equation}
In the above expression, $\Lambda_0 $ is an upper bound for the cutoff in
our model of the VHS, that we take as 1 eV. The other important factor
in the resolution of (\ref{erg}) is the choice of initial conditions for
$V_{2,2}$ and $V_{2,4}$ at the upper cutoff $\Lambda_0 $. In this respect,
we have taken a value of $U = 4$ eV for the bare on-site repulsion,
which is between the estimates made for graphite and carbon nanotubes\cite{perf}. 
In order to go beyond the perturbative particle-hole corrections to the 
BCS vertex, we have summed up the series of leading logarithms obtained by
iteration of the particle-hole susceptibilities in the diagrams of Fig. 
\ref{two}. Thus, we have constrained the initial couplings by the 
condition
\begin{equation}
3 (V_{2,2} + 2 V_{2,4})  =   \frac{U}{1 - 3 U \chi ({\bf 0}, \mu )}
                    - \frac{U}{1 - 2 U \chi ({\bf Q}, \mu )}
\end{equation}
The precise values of $V_{2,2}$ and $V_{2,4}$ have been obtained by adding
the other constraint mentioned below Eq. (\ref{cond2}), which reads
$3 (2 V_{2,4} - V_{2,2})  = 
\widetilde{V} (\pi /6, -\pi /6 ) - \widetilde{V} (\pi /2, -\pi /2 )$.
In these conditions, we have determined the point of the pairing
instability in terms of the energy $\Lambda $ at which the solution of
Eq. (\ref{erg}) diverges, marking the position of a pole in the 
BCS vertex. The results are shown in Fig. \ref{three}, where we have
represented the critical point in a temperature scale after trading the
energy variable $\Lambda $ by the thermal energy $k_B T$.

We observe that the scale of the pairing instability depends drastically
on the value of the chemical potential $\mu $ measuring the deviation of 
the VHS from the Fermi level. The plot of Fig. \ref{three} shows anyhow that 
the instability exists irrespective of the value of $\mu $. 
We recall in this respect that the Kohn-Luttinger mechanism was proposed
to put forward the idea that any Fermi liquid is unstable at 
sufficiently low temperature\cite{kl}. 
On the other hand, an inflection point can be seen in the plot of Fig. 
\ref{three} for a value of $\mu $ slightly below 0.2 meV. That feature 
corresponds to the case in which the energy scale of the instability (and 
the scale of the gap) coincides with the deviation $\mu $ of the VHS from the 
Fermi level. Values of $\mu $ to the left of the inflection point correspond 
therefore to the regime where the pairing instability is driven all the way 
by the VHS.

The possibility of finding a pairing instability in graphene at temperatures
of the order of $\sim 10$ K may rely on the ability to make a fine tuning of
the Fermi level to the VHS. This may be feasible as long as
the proximity to the divergent density of states corresponds to a situation 
with a very large compressibility, which is energetically very favorable. 
It has been actually shown that the VHS may pin the Fermi level over a range
of doping levels, forcing the breakdown of the uniform charge distribution into 
patches with uneven electron density (phase separation) so that the Van Hove 
filling is reached in one of the phases\cite{ch}. 

In conclusion, we have seen that placing graphene in the proximity of the
VHS of its valence band may be a good instance to induce
a superconducting instability in the electron system. The origin of this
effect lies on the large anisotropy of the $e$-$e$ scattering along the 
Fermi line, which leads to an attractive coupling in a channel with 
$d$-wave symmetry. We have shown that the scale of the pairing instability
may be pushed up to temperatures larger than 10 K, depending on the ability
to tune the system to the proximity of the VHS.

The financial support of the Ministerio de Educaci\'on y Ciencia
(Spain) through grant FIS2005-05478-C02-02 is gratefully
acknowledged.

\end{document}